\documentclass[preprint,authoryear,11pt]{elsarticle_mod}
\pdfoutput=1

\usepackage{SIunits}

\usepackage{placeins}

\begin{document}

\begin{frontmatter}



\title{Tensile material properties of human rib cortical bone under quasi-static and dynamic failure loading and influence of the bone microstucture on failure characteristics}

\author[uva]{Damien Subit\corref{cor}}\ead{subit@virginia.edu}\ead[url]{www.centerforappliedbiomechanics.org}
\author[ecip]{Eduardo del Pozo de Dios}\ead{edelpozod@unav.es}\ead[url]{www.unav.es/ecip}
\author[Juan]{Juan Vel\'azquez-Ameijide}\ead{juan.velazquez@upc.edu}\ead[url]{www.upc.edu}
\author[uva,unav]{Carlos Arregui-Dalmases}\ead{carlosarregui@unav.es}\ead[url]{www.unav.es/ecip}
\author[uva]{Jeff Crandall}\ead{jrc2h@virginia.edu}\ead[url]{www.centerforappliedbiomechanics.org}
\cortext[cor]{Corresponding author}

\address[uva]{University of Virginia, Center for Applied Biomechanics, 4040 Lewis and Clark drive, Charlottesville, VA 22902, USA}
\address[ecip]{Universidad de Navarra, European Center for Injury Prevention, Irunlarrea 1 (ed. Los Casta\~nos), 31008 Pamplona, Navarra, Spain}
\address[Juan]{Universitat Polit\`ecnica de Catalunya, Dept. Resist\`encia de Materials i Estructures a l'Enginyeria, Comte d'Urgell 187, 08036 Barcelona, Spain}
\address[unav]{Universidad de Navarra, Preventive and Public Health Dpt., School of Medicine, Irunlarrea 1, Ed. investigaci\'on (despacho 2440), 31008 Pamplona, Navarra, Spain}

\begin{abstract}
Finite element models of the thorax are being developed to assist engineers and vehicle safety researchers with the design and validation of countermeasures such as advanced restrain systems.  Computational models have become more refined with increasing geometrical complexity as element size decreases. These finite element models can now capture small geometrical features with an attempt to predict fracture. However, the bone material properties currently available, and in particular the rate sensitivity, have been mainly determined from compression tests or tests on long bones. There is a need for a new set of material properties for the human rib cortical  bone. With this objective, a new clamping technique was developed to test small bone coupons under tensile loading. This technique allows for applying minimal constraints to the coupon during clamping and ensures that the main type of loading is tension.
Ten coupons were harvested from the cortical shell of the sixth and seventh left ribs from three cadavers. The coupons were tested to fracture under quasi-static (target strain rate of rate of 0.07 \%/s) and dynamic loading (target strain rate of 170 \%/s). Prior to testing, each coupon was imaged with a computed micro-tomograph to document the bone microstructure. An optical method was used to determine the strain field in the coupon for the quasi-static tests. The rib bone coupons were found to be elastic, with brittle fracture. No plastic behavior was observed in this test series. The bone coupons were assumed isotropic, homogeneous and elastic linear, and the average Young's modulus for the quasi-static tests (13.5 GPa) and the failure stresses (quasi-static: 112 MPa, dynamic: 124.6 MPa) were in line with published data.  Fracture however did not always occur in the gage area where the cross-sectional area was the smallest, which contradicted the assumption of isotropy and homogeneity. The comparison of the results obtained in the current study with published results on tibia and femur bone coupons suggests that the effective cross-section has an effect on the calculated material properties, and that further analysis of the bone microstructure is required to establish  the rib cortical bone fracture mechanism.
\end{abstract}


\end{frontmatter}

\section{Introduction}
\begin{figure}[!h]
\begin{center}
  \includegraphics[width=35mm]{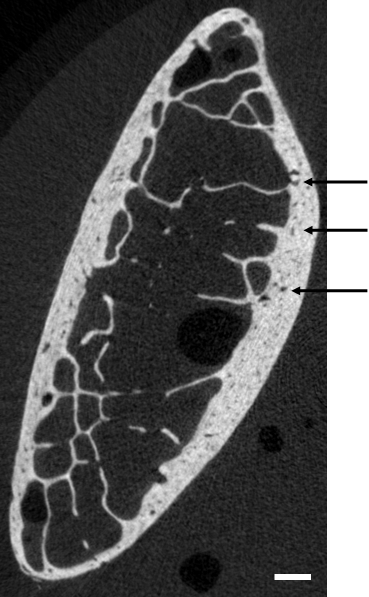}\\
  \caption{Cross-section of the left 6\textsuperscript{th} from the antero-lateral aspect, from micro-computed tomography. The arrows indicate voids. The bar scale is 1 mm long.}\label{fig:cross_section}
  \end{center}
\end{figure}

Rib fractures are a good indicator of the severity of an impact to the thorax as the protection to the internal organs such as the lungs and the heart is greatly reduced with the increasing number of fractured ribs (abbreviated injury scale, \cite{AAAM1995}). Injury mechanisms for the ribs and the whole rib cage have been widely studied, either through experiments \citep{Kent2004, Vezin2009, Kuppa1998, Trosseille2008, Hallman2010, Petitjean2002, Lessley2010ijc} or computational simulations \citep{Murakami2006, Lizee1998, Song2009, Robin2001, Vezin2005, Shigeta2009, Kimpara2005, Plank1989, Ruan2003, Li2010job, Li2010mep,Kimpara2006} to determine injury mechanisms and thresholds under diverse load conditions. A significant milestone was achieved in the characterization of the strength of the thorax by accounting for the geometrical variations in the rib cage and the rib themselves, and for the effects of biological variations such as aging \citep{Berthet2005,Ito2009,Gayzik2008,Kent2004}. Ribs were shown to have a complex geometry that includes variation in the shape of the cross-section along the rib axis \citep{Kindig2009}, an increase of the twist from the posterior to the anterior aspect \citep{Mohr2007}, as well as a non uniform distribution of the cortical thickness \citep{Choi2009}. In a recent study, \citet{Li2010mep} investigated the sensitivity of the rib structural response obtained from a computational model to (a) the accuracy in the reconstruction of the rib (quantified by the mesh density), (b) the cortical thickness distribution and (c) the material properties. The response of the rib finite element model was found to be little sensitive to the choice of material properties, whereas clear trends were observed in the effect of the mesh density and cortical thickness distribution. In finite element modeling, the cortical shell is defined as a continuum made of pure cortical bone \citep{Li2011}. However, rib cortical bone has voids and is not homogeneous (figure \ref{fig:cross_section}), and therefore the bone material properties documented for homogeneous bone are likely to inadequately represent the mechanical behavior of rib cortical bone.

Bone tensile material properties have been reported extensively for bovine bones \citep{Pithioux2004,Ferreira2006,Adharapurapu2006,Saha1974,Wright1976,Wright1977,Crowninshield1974}, but less often for human bone \citep{Saha1976,Kemper2005,Keller1990,Hansen2008}. The experimental tests by \citet{Kemper2005} are the only study targeted at human rib bones. Delicate "dog bone" coupons (length: \unit{30}{\milli\meter}, width: \unit{9}{\milli\meter} at the ends, \unit{2.5}{\milli\meter} in the gage area, thickness: \unit{0.5}{\milli\meter}) were machined from cortical rib bones, and a slack adaptor system was designed to apply a constant strain rate, with a target of 50\%/s. Special care was taken to prevent the breaking of the samples during handling. However, for more than 80 \% of the tests (95 out of 117 tests) the bone samples fractured in the grip area. For the other tests, the actual fracture location was not reported and therefore one can only assume that the fracture occurred in the gage area. Dog bone samples have a weak point by design -- the gage area where the width is much smaller than that at the ends of the sample -- so as to generate greater tensile stress in the gage area and ensure that fracture is caused by tension, at least under the assumption that the sample is homogeneous. The fact that fracture occurred in the grip area so often suggests either that the test apparatus caused the fracture, for instance because of inadequate clamping force or misalignment of the top and bottom grips, or that the hypothesis of homogeneity did not hold.
In addition, \citet{Kemper2005} used an extensometer to measure the average strain for the gage area that may have interfered with the mechanical response of the samples. The questions raised in \citet{Kemper2005} and the paucity of rib bone material data highlighted the need for new experimental data.
The goal of the current study was to perform new tensile tests of rib bone coupons to link the fracture characteristics to the coupon bone microstructure. To do so, an experimental protocol was developed to machine and test rib cortical bone coupons of constant thickness under quasi-static and dynamic loading. All the samples were imaged prior to testing using a micro-computed tomograph. This protocol was evaluated on coupons machined from the left sixth and seventh ribs harvested from three post mortem human subjects.

\section{Materials and methods}
\subsection{Coupon preparation}
\begin{figure}[!h]
\centering
  \includegraphics[width=130mm]{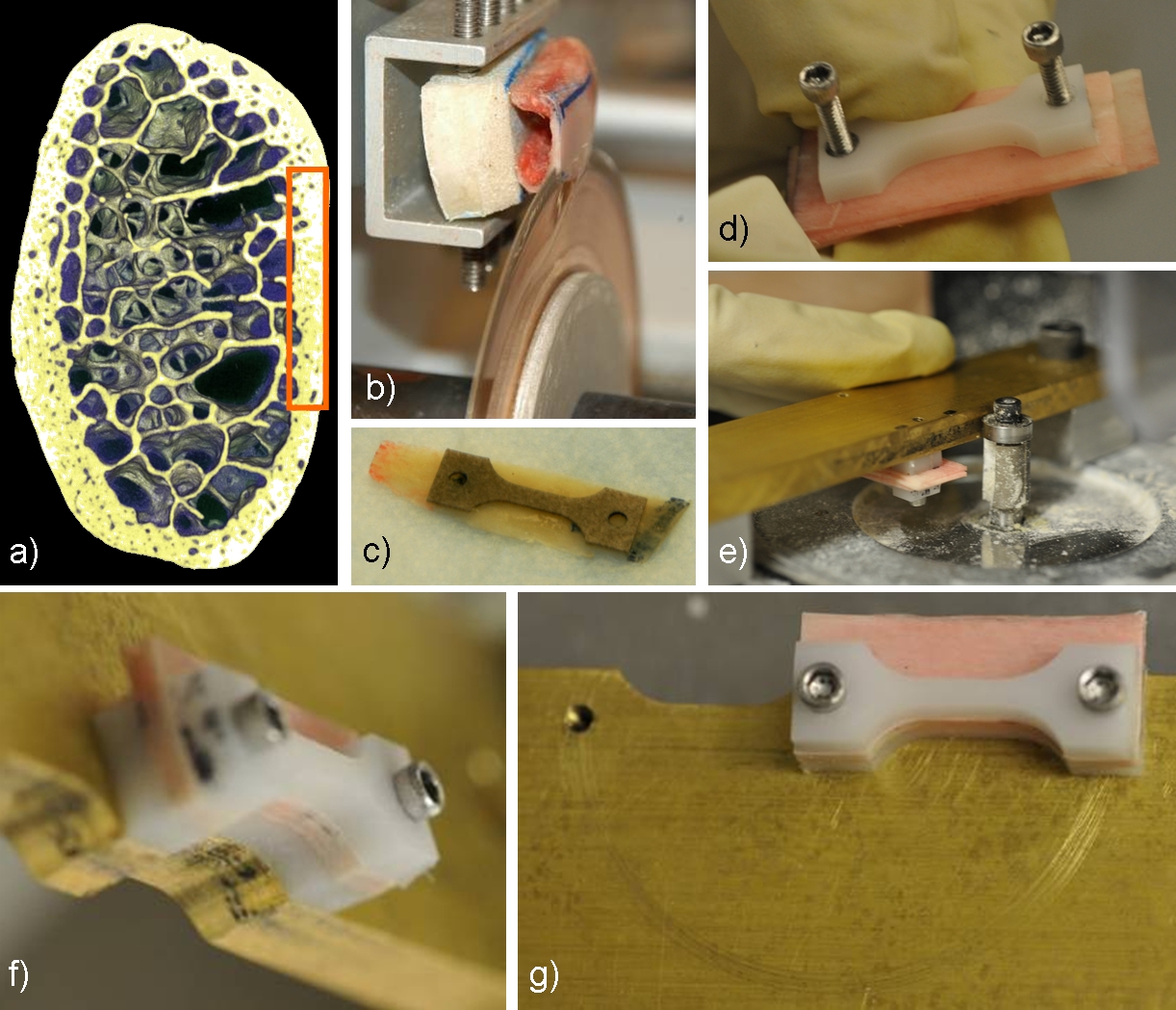}\\
  \caption{Machining of bone coupons.}\label{fig:machining}
\end{figure}

\begin{figure}[!h]
  \centering
  \includegraphics[width=70mm]{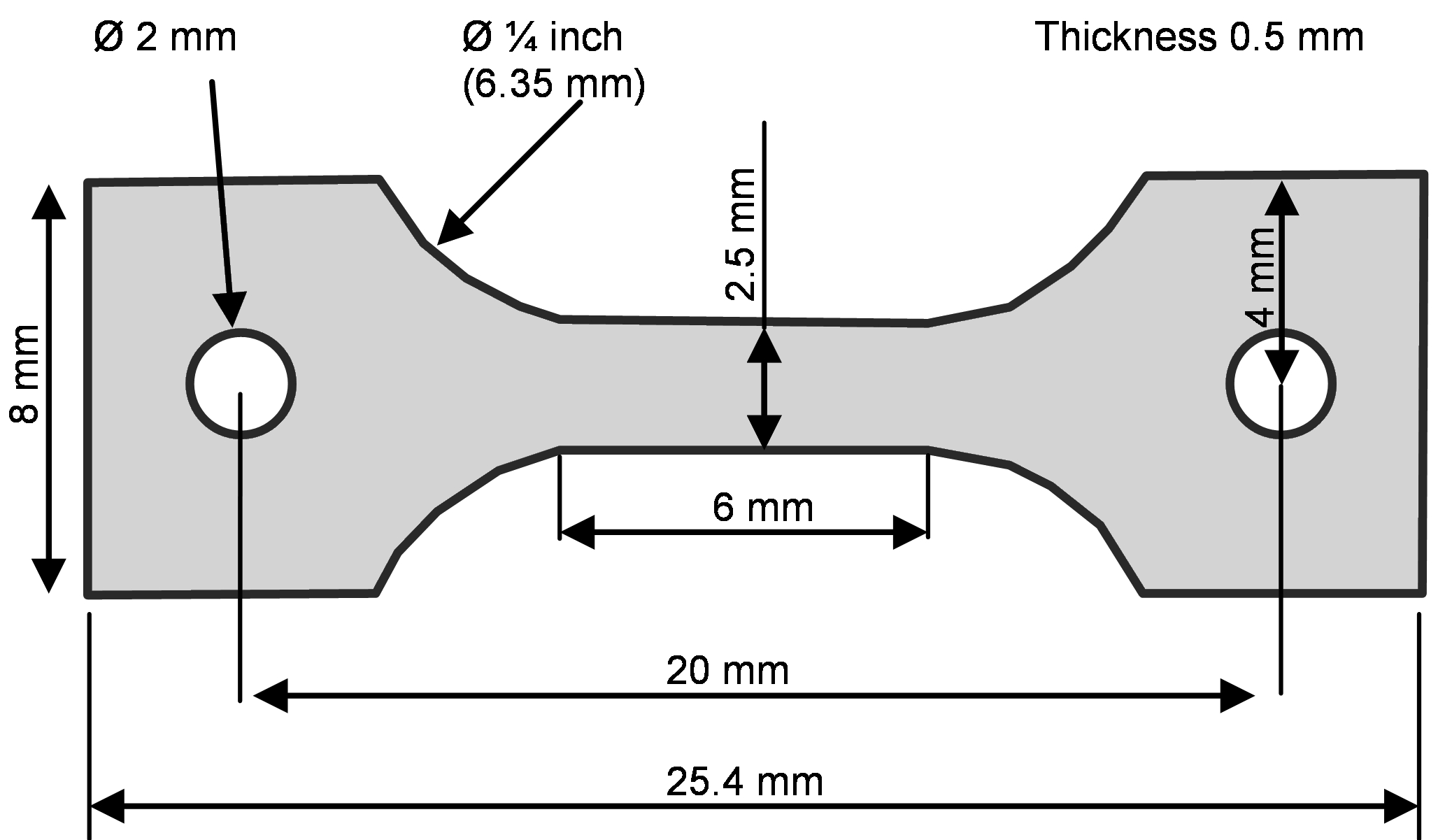}\\
  \caption{Dimensions of the bone coupons. The 2 holes are used only for references for machining and testing, and do not carry any load.}\label{fig:coupon_dimensions}
\end{figure}

The ribs were first cut into five 60-mm long pieces named A (most anterior piece) to E (most posterior piece). Isotropic images similar to figure \ref{fig:cross_section} were obtained with a micro-computed tomograph (Scanco Medical Viva CT40, Br\"uttisellen, Switzerland) at a resolution of \unit{30}{\micro\meter}/pixel and used to determine where the bone coupons could be machined based on the curvature and the thickness of the cortical shell (Figure \ref{fig:machining}-a). Next, the periosteum was removed, and the sample was shortened with a small abrasive band saw equipped with a diamond blade (Diamond bone band saw, Mar-med inc., Cleveland, OH, USA). The next cuts were performed with a low speed diamond saw (IsoMet low speed saw, Buehler Ltd, Lake Bluff, IL, USA, equipped with a 15HC Diamond blade) that can execute parallel cuts with micrometric accuracy. The sample was glued to a piece of hard foam that was then clamped in the sample holder in the desired orientation thanks to screws with spear tips (Figure \ref{fig:machining}-b). This step was critical in the machining process: the orientation of the bone slab had to be adjusted so that the cut performed with the low speed diamond saw would be entirely in the cortical shell. A first cut was made to generate the outermost flat surface. From there, an other cut was made parallel to the first one to create a flat piece of cortical bone (referred to as wafer) that was \unit{0.5}{\milli\meter} thick (Figure \ref{fig:machining}-c). Next the two holes used to reference the position of the bone coupon (figure \ref{fig:coupon_dimensions}) were drilled in the wafer.

The next step consisted in machining the wafer to give it its final shape. This happened in two successive actions. First, the wafer was shortened to the desired length, and second, the narrow section of the coupon was machined. These two actions utilized the same principle: the wafer was sandwiched between two pieces of plastic (Figure \ref{fig:machining}-d), and attached thanks to two screws to a template made of brass. A router equipped with a bearing flush trim bit was used to trim the bone wafer to the desired shape: the bearing was kept in contact with the brass template that was used as a guide (Figure \ref{fig:machining}-e). To make the narrow section of the coupon, a template with two grooves was used. After the first side was machined, the sample was rotated about one screw and set along the other groove to machine the other side (Figures \ref{fig:machining}-f and g). During the entire procedure, the orientation of the wafer relative to the original bone was tracked and the bone was kept hydrated (0.9 \% saline solution). The coupons were then stored in a tube filled with saline solution and kept in a fridge until the test day (between 1 and 4 days after preparation) to ensure proper conservation, and imaged with the micro-tomograph. The thickness and width at the center of each coupon were measured prior to each test.

\subsection{Test fixture and procedure}
\begin{figure}[!h]
\centering
\includegraphics[width=30mm]{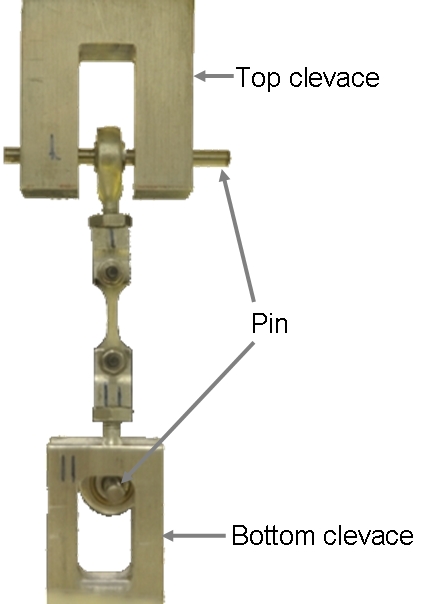}
\caption{Clamping system for the coupons.}\label{fig:clamping_system}
\end{figure}

\begin{figure}[!h]
  \centering
  \includegraphics[width=50mm]{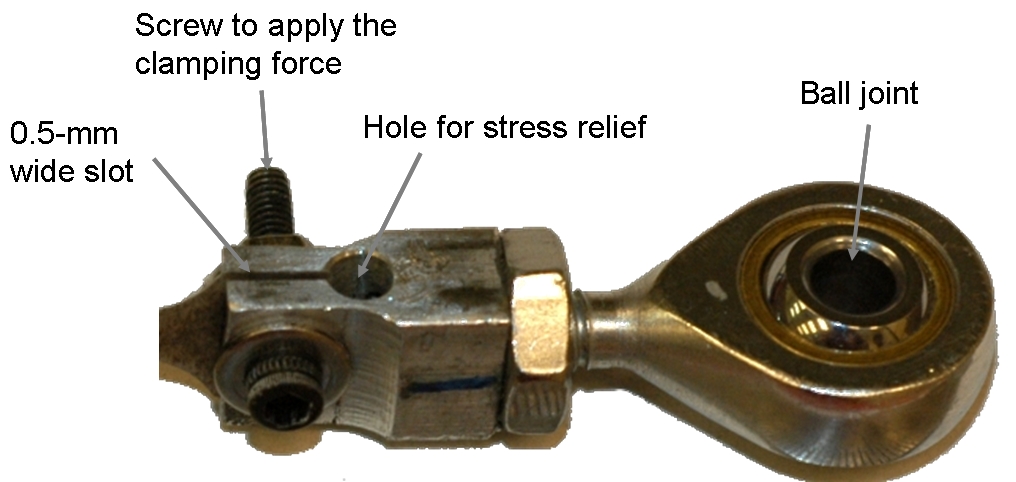}\\
  \caption{Close-up view of the clamp.}\label{fig:clamp}
\end{figure}

A hydraulic tensile machine was used (Model 8874, Instron Inc, Norwood, MA, USA). A clamping system made of aluminum was designed to avoid misalignment between the top and the bottom ends of the coupon (figure \ref{fig:clamping_system}). Because of their fragility, it was critical to be able to install the coupons on the test machine without applying any load that could generate bending, shear or too much tension. Conventional fixed wedge clamps would not ensure the required control of the load during the installation of the coupon. Therefore small low-mass clamps (10.9 grams each) were designed to be attached to the coupon (figure \ref{fig:clamp}) prior to mounting the sample to the test machine. A torque of \unit{1}{Nm} was applied to the screws used to clamp the samples. Preliminary tests showed that this torque value was adequate to prevent slippage of the sample, while avoiding bone crushing.

Pins that go through the rod-end ball joints were used to connect the clamps to the slitted blocks affixed to the base of the tensile machine (bottom clamp) and to the end of the piston (top clamp, Figure \ref{fig:clamping_system}).

The top clamp was first installed and the coupon equipped with the clamps was let to hang. The position of the sample was adjusted by moving the piston so that the bottom pin could go through the bottom ball joint without applying any load.
The piston was then slowly moved up until the bottom pin touches the clevace, resulting in a preload of about \unit{2}{N}. The bone coupons were stored in saline solution until the paint pattern was applied, a few minutes prior to testing (see section \ref{ssec:data_acquisition}).

\subsection{Test matrix}
The sixth and seventh right ribs were harvested from the three subjects included in \citet{Lessley2010} (Table \ref{tab:pmhs_info}). Ten bone coupons were machined, and tested up to fracture with a constant velocity displacement under quasi static (\unit{0.01}{mm/s} and \unit{0.02}{mm/s}) or dynamic (\unit{24}{mm/s}) rates (Table \ref{tab:test_matrix}). The velocity of the applied displacement was determined to generate a strain rate similar to that reported on in \citet{Lessley2010} for the dynamic tests.

\begin{table}[!h]
  \centering
  \begin{tabular}{ccccc}
    \hline
    Subject &	Age at time of death & Cause of death    & Body mass (kg) & Stature (cm)\\
    \hline
    468  &  67                   &	Stroke             &  64              & 166         \\
    473	 &	54                   &	Brain aneurysm     &  73              & 182       \\
    480	 &	71                   &	Laryngeal cancer   &  70            & 182         \\
    \hline
  \end{tabular}
  \caption{Subjects characteristics.}\label{tab:pmhs_info}
\end{table}

\begin{table}[!h]
  \centering
  \begin{tabular}{cccccc}
    \hline
Subject & Rib level & Location & Aspect & Velocity (mm/s) & Strain rate (\%/s)\\
\hline
    468 & 6 & A & Lateral & 0.02 & 0.089\\
        & 7 & B & Lateral & 0.01 & 0.055\\
        & 7 & C & Lateral & 24   & 170\\
        \hline
    473 & 6 & B & Medial  & 0.01 & 0.046\\
        & 7 & B & Lateral & 0.01 & 0.043 \\
        & 7 & B & Medial  & 24   & 170\\
        & 7 & C & Lateral & 24   & 170\\
        \hline
    480 & 6 & A & Lateral & 0.01 & 0.069\\
        & 7 & B & Lateral & 24   & 170\\
        & 7 & C & Lateral & 0.01 & 0.055\\
        \hline
  \end{tabular}
  \caption{Test matrix. The actual strain rates are provided for the quasi-static tests, whereas only the target strain rate is provided for the dynamic tests as the strain time history data could not be determined (see section \ref{ssec:data_acquisition}).}\label{tab:test_matrix}
\end{table}

\subsection{Data acquisition and processing}\label{ssec:data_acquisition}
The tensile load was measured by a three axis loadcell located under the bottom clamp (model 6085, Denton Inc, Plymouth, MI, USA), connected to a standard data acquisition system (DEWE-2010, Dewetron GmbH, Graz, Austria). The tensile and shear loads were measured.  The sampling rate was \unit{500}{Hz} for the quasi-static tests, and \unit{100}{kHz} for the dynamic tests, and the signals were filtered using a low-pass 2\textsuperscript{nd} Butterworth filter with a cut-off frequency of \unit{10}{Hz} for the quasi-static tests and \unit{600}{Hz} for the dynamic tests. These filters were chosen to remove the noise of the signals while keeping their shapes and introducing negligeable time-shift.

\begin{figure}[!h]
  \centering
  \includegraphics[width=20mm]{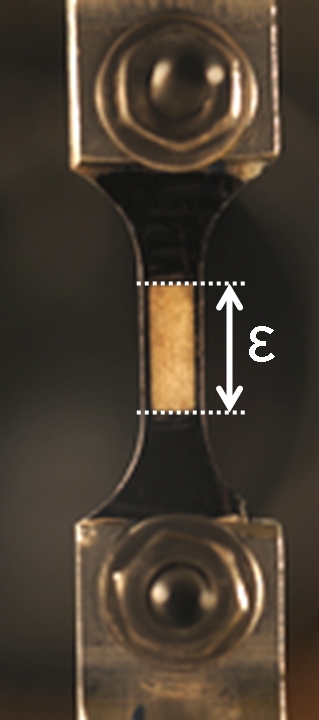}\\
  \caption{Coupon with high contrast pattern prior to test.}\label{fig:coupon_strain}
\end{figure}

For the quasi-static tests, the strain was measured by performing image analysis of pictures taken during the tests with a 12-megapixel single-lens reflex camera equipped with a 100-mm macro lens and triggered by a programmable controller (The Time Machine, Mumford Micro Systems, Santa Barbara, CA, USA) that allows for tripping the camera shutter at a constant frequency of 3 frames per second. The controller signal was also sampled by the data acquisition system to synchronize strain and force measurements. The outermost surface of the coupon (with respect to its orientation relative to the rib before machining) was painted with a black and white pattern (figure \ref{fig:coupon_strain}) to allow for strain measurement based on contrast analysis \citep{Frank2008}. Strain could not be measured for dynamic tests due to the unavailability of a continuous (non-flickering) light source.
The tensile stress was defined as the ratio of the tensile force by the cross-sectional area in the middle of the coupon. The effective Young's modulus (referred to as Young's modulus) was defined as the slope of the strain-stress curve between 0 and \unit{0.5}{\%}. For the quasi-static tests, stress and strain could be measured. For the dynamic tests, only failure stress could be determined. In addition, fracture location was documented by measuring the distance between the fracture line and the anterior end of the coupon.

\subsection{Analysis of the images from micro-computed tomography}
The images obtained with the micro-tomograph (DICOM format) were analyzed to determine the actual cross-sectional area that accounts for the presence of voids in the cortical microstructure and for the cross-sections not being perfectly rectangular. A Matlab script (2010a, The Mathworks, Natick, MA, USA) was written to perform the segmentation of the DICOM images: a threshold of 3000 Housfield Unit (HU) was used, and all the pixels with a HU value greater than the threshold were considered as bone. The position along the coupon was normalized, with 0 and 1 being respectively the anterior and posterior aspects of the coupon (relative to the rib).

\section{Results}

\subsection{Material properties}
\begin{figure}[!h]
  \centering
  \includegraphics[width=70mm]{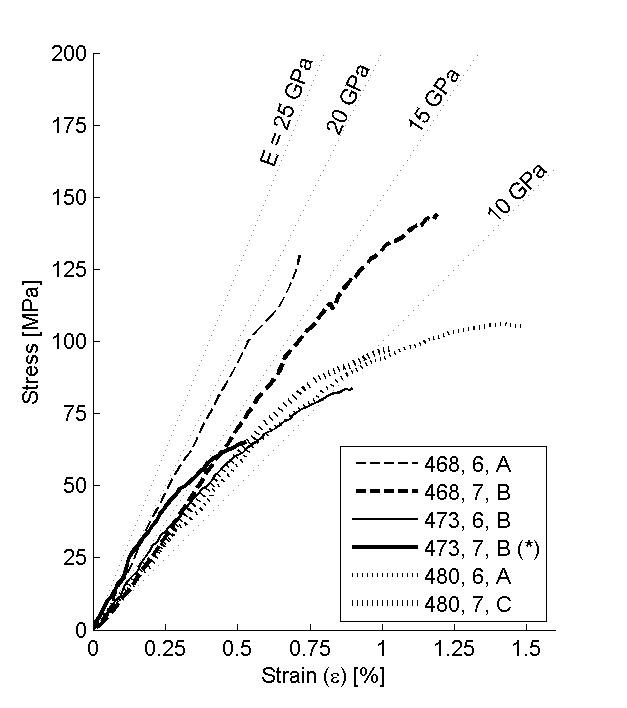}\\
  \caption{Strain-stress curves for the quasi-static tests. *: sample fractured in the clamp.}\label{fig:qs_curve}
\end{figure}

Strain-stress curves were plotted for all the quasi-static tests (figure \ref{fig:qs_curve}). The material properties were determined by assuming a linear elastic model between 0 and \unit{0.5}{\%} of tensile strain (table \ref{tab:mat_prop}). Young's modulus ranged from 11.4 to \unit{18.5}{GPa}, failure stress from 83.4 to \unit{143.9}{MPa}, and failure strain from 0.71 to \unit{1.49}{\%} in quasi-static, and failure stress ranged from 94.7 to \unit{155.9}{MPa} in dynamic.

\begin{table}[!h]
  \centering
  \begin{tabular}{ccccl}
    \hline
    Coupon & E (GPa) & $\sigma_{failure}$ (MPa) & $\varepsilon_{failure}$ (\%) & Fracture location\\
    \hline
    Quasi-static & & & & \\
    \hline
    468, 6, A & 18.5 & 129.9 & 0.71 & \phantom{qq}\includegraphics[width=2cm]{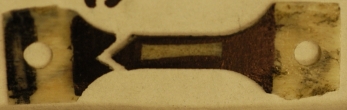}\\ 
    468, 7, B & 14 & 143.9 & 1.19 & \phantom{qq}\includegraphics[width=2cm]{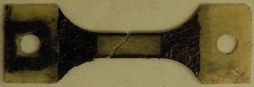}\\ 
    473, 6, B & 12.2 & 83.4  & 0.9  & \phantom{qq}\includegraphics[width=2cm]{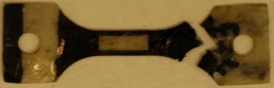}\\ 
    473, 7, B & 12.8 & (*)   & (*)  & \phantom{qq}\includegraphics[width=2cm]{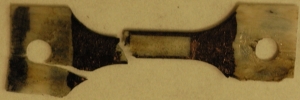}\\ 
    480, 6, A & 11.4   & 106   & 1.49 & \phantom{qq}\includegraphics[width=2cm]{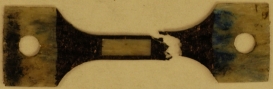}\\ 
    480, 7, C & 12.2   & 97.3  & 1.02 & \phantom{qq}\includegraphics[width=2cm]{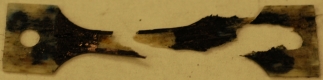}$\dag$\\ 
    Mean values  & 13.5 $\pm$ 2.6 & 112.1 $\pm$ 24.5 & 1.06 $\pm$ 0.29\\
    $\pm$ standard deviation & & & \\
    \hline
    Dynamic & & & &\\
    \hline
    468, 7, C & & 140.8 & & \phantom{qq}\includegraphics[width=2cm]{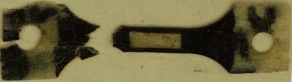}\\
    473, 7, B & & 155.9 & & \phantom{qq}\includegraphics[width=2cm]{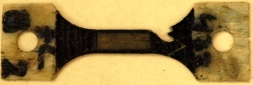}\\
    473, 7, C & & 107   & & \phantom{qq}\includegraphics[width=2cm]{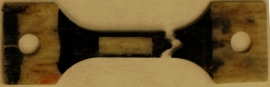}\\
    480, 7, B & & 94.7  & & \phantom{qq}\includegraphics[width=2cm]{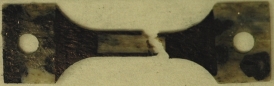}\\
    Mean values  & & 124.6 $\pm$ 28.5 & \\
    $\pm$ standard deviation & & & \\
    \hline
  \end{tabular}
  \caption{Rib cortical bone material properties for the quasi-static and dynamic tests, and images of the fractured coupons (posterior surface). The anterior end is the left side. (*) sample fractured in the clamp. ($\dag$) the fracture by the right hole occurred after the test during handling.}\label{tab:mat_prop}
\end{table}

\subsection{Fracture location}
\begin{figure}[!h]
\hspace*{-1.5cm}
   \includegraphics[width=150mm]{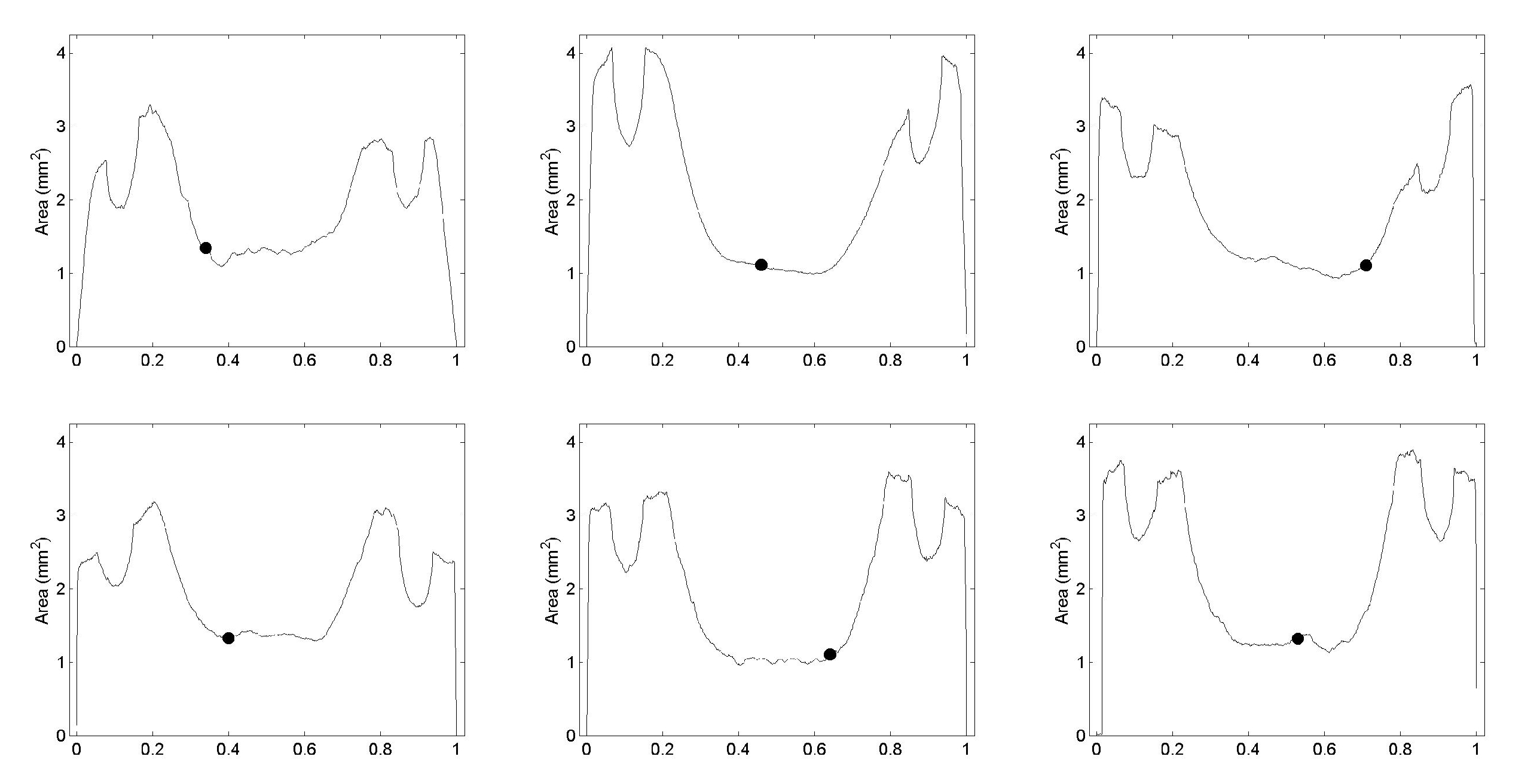}
    \caption{Variation of the cross-section (quasi-static tests). The fracture location is indicated by the dot. Zero is the anterior end of the coupon, and one is its posterior end.}\label{fig:cross_section_change_qs}
\end{figure}

\begin{figure}[!h]
\begin{center}
   \includegraphics[width=100mm]{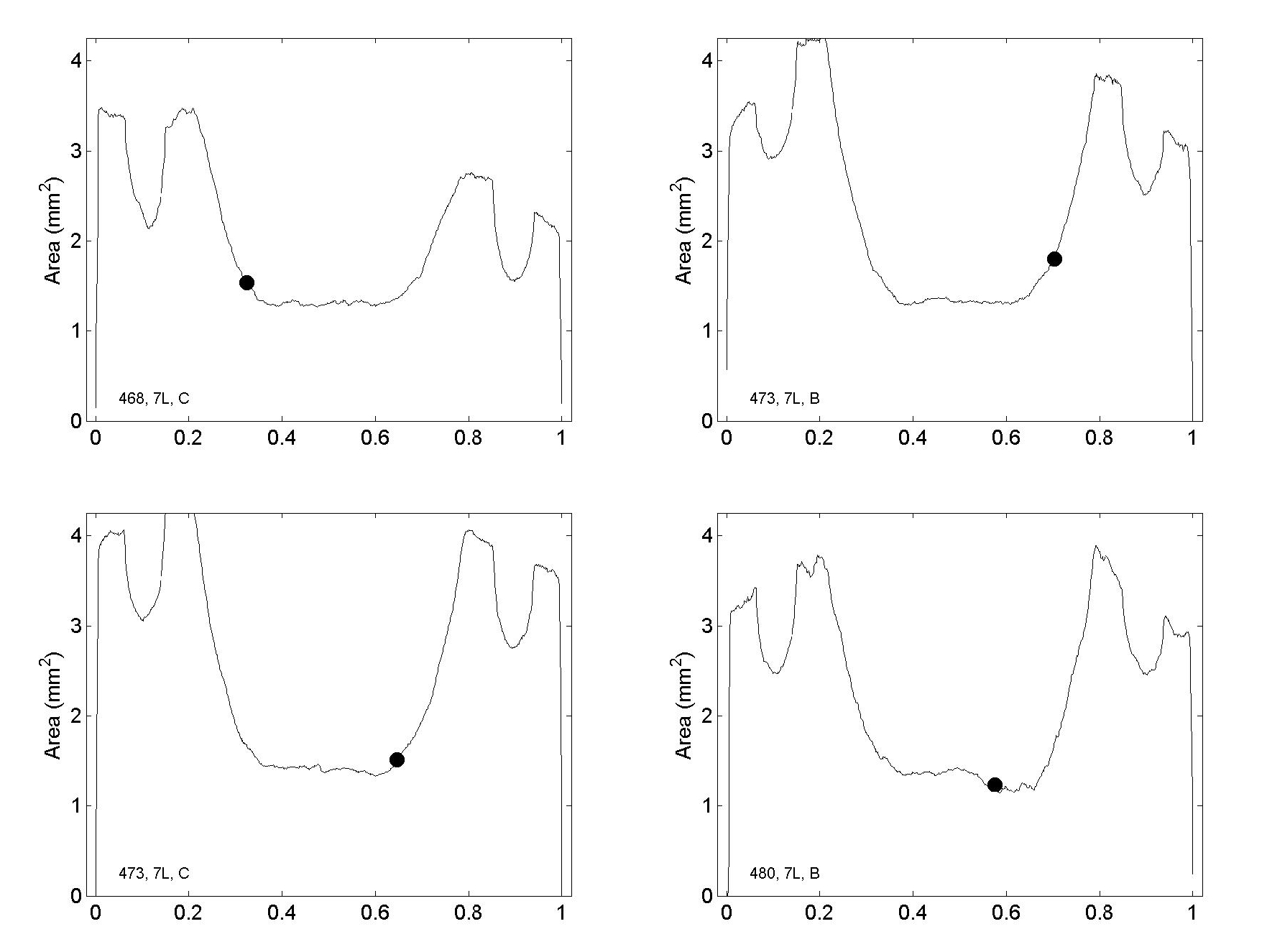}
    \caption{Variation of the cross-section (dynamic tests). The fracture location is indicated by the dot. Zero is the anterior end of the coupon, and 1 is its posterior end.}\label{fig:cross_section_change_dyn}
\end{center}
\end{figure}

Fracture occurred outside the grip area in nine of the ten tests. Fracture occurred first within the clamping area for one sample (473, 7, B) because the sample slid of the clamp, causing the screw to bear the tension load. The posterior end (right, figure \ref{tab:mat_prop}) for this sample was too thin (about \unit{0.43}{mm}) to provide sufficient clamping load, and the sample slid relative to the clamp after a certain tensile load was applied. The analysis of the images collected during the test for strain measurement confirms that sliding occurred after \unit{0.5}{\%} deformation; therefore only the Young's modulus could be determined for this test. The variation of the cross-section along the length of the coupons shows that the gage area cross-sectional areas was close to the target (\unit{1.25}{mm}) for all the coupons (figures \ref{fig:cross_section_change_qs} and \ref{fig:cross_section_change_dyn}), whereas the cross-sectional area of the ends varied greatly from sample to sample. The images obtained from micro-computed tomography are provided for each coupon (\ref{app1}).

\section{Discussion}

\subsection{Material properties}
The rib bone mechanical properties were calculated based on the assumption that the rib cortical bone was homogeneous and isotropic. The average value and standard deviation for the Young's moduli determined for the quasi-static tests (13.5 $\pm$ 2.6 GPa) were similar to the 13.9 $\pm$ 3.7 GPa reported in \citet{Kemper2005}. The failure stresses were also in the same range for both the quasi-static (121 $\pm$ 24.5 MPa) and dynamic tests (124.6 $\pm$ 28.5 MPa) in the current study compared to \citeauthor{Kemper2005} (124.3 $\pm$ 35.4 MPa). Although the target strain rates were different in these two studies (\unit{0.07}{\%/s} for the quasi-static and \unit{170}{\%/s} for the dynamic tests in the current study, \unit{50}{\%/s} in \citeauthor{Kemper2005}), the measured strain rates are actually wide spread (\unit{0.043}{\%/s} to \unit{0.089}{\%/s} for the current study, and from \unit{9}{\%/s} to \unit{90}{\%/s} in \citeauthor{Kemper2005}). Based on \citet{Hansen2008} analysis of several studies reporting bone material data, the Young's modulus is little rate sensitive for strain rates below \unit{200}{\%/s}. Therefore the comparison of the results reported in the current study with that reported in \citeauthor{Kemper2005} is justified. A major difference between these two studies is the non linearity, as \citeauthor{Kemper2005} reported on a substantial yield behavior, and much greater failure strains (2.68 $\pm$ 1.4 \%) than what was determined here (1.05 $\pm$ 0.29 \%). In addition, the failure strains reported on in \citeauthor{Kemper2005} range from 0.52 to 6.56\%/s.
This range is high compared to what is commonly used for bone. It is not clear whether the discrepancies are artifacts because of the experimental methods or due to age differences, as the subjects in the current study are a little bit older than in \citet{Kemper2005}. The coupons machined in \citeauthor{Kemper2005} had a wide range of cross-sectional areas (from 0.38 to 2.08 mm\textsuperscript{2}), whereas the cross-sectional areas of coupons used for the present tests were more uniform (1.25  mm\textsuperscript{2}). The effect of the size of the coupons on the measured mechanical properties is not well understood, in particular how the loading varies with the varying cross-sectional area. In the analysis performed in the two studies, tensile is assumed to be the main type of loading. Thanks to the image analysis performed to measure strain in the current study, it was possible to ensure that there was no slipping in the clamps (except for the one test identified in the results section), whereas \citeauthor{Kemper2005} reported that most of the coupons broke in the clamp area which suggests that these fractures were artifactual of the experimental set-up. Too few samples were tested in this study to determine whether bone properties vary as a function of the location along the rib.

The failure stress was found to be on average higher for the dynamic tests than for the quasi-static tests (however fewer samples were tested under dynamic loading). There is very little data in the published literature regarding rib cortical bone rate sensitivity under tensile loading. The data reported on in \citet{McElhaney1966} are often mistakenly used to describe the rate sensitivity of the bone material properties in tension, although the results established by \citeauthor{McElhaney1966} are based on compression tests. Other studies such as \citet{Crowninshield1974} used animal tissue and the bone samples were machined in the lower extremity; consequently the coupons are bigger. In addition, the mechanical function of a a bovine leg is different from that of a human rib, and therefore the microstructure is likely to be different. It is indeed true that the coupons machined in the human cortical femur and tibia bone are denser than that machined in the rib bone (on-going analysis of unpublished test data). Therefore, the effective modulus is higher for the lower extremity coupons than for the rib ones (table \ref{tab:summary_mat_prop}).

The coupons machined in the human cortical bone are heterogeneous (\ref{app1}), and the actual cross-sections (determined by taking into account the presence of the voids) were smaller than the target of \unit{1.25}{mm\textsuperscript{2}} (figures \ref{fig:cross_section_change_qs} and \ref{fig:cross_section_change_dyn}). When compared to the material parameters determined for the human femur (table \ref{tab:summary_mat_prop}), rib bone exhibits a lower Young's modulus, but the failure stress and failure strain are in the same range. Although the strains were not measured using the same methods for all these studies (extensometer or non contact optical measurements), it makes sense that the effective modulus appears smaller for the rib bones. The voids in the rib bone microstructure are not accounted for in the measurement of the cross-sectional area as the coupon cross-section is assumed to be rectangular and its area is approximated by measuring its thickness and width. With the femur cortical bone being denser than the rib cortical bone, coupons made of femur bone or rib bone with the same external dimensions would appear to have the same cross-sectional area, whereas the actual cross-section of the rib bone coupon will be smaller because of the voids.

\begin{table}
{\footnotesize
\hspace*{-3cm}
\begin{tabular}{lcccc}
\hline
Study & Bone type & Young's modulus (GPa) & Failure stress (MPa) & Failure strain (\%) \\
\hline
\citet{Yamada1970} - Dry  & Human long bones  & 18.3           & 140              & 1.49    \\
\citet{Yamada1970} - Wet  & Human long bones  & 21.1           & 172              & 1.3      \\
\citet{Hansen2008}        & Human femora      & 16.1 $\pm$ 2.1 & 119.8 $\pm$ 20.7 & 2.5  $\pm$ 0.8 \\
\citet{Kemper2005}        & Human ribs        & 13.9 $\pm$ 3.7 & 124.3 $\pm$ 35.4 & 2.68 $\pm$ 1.4 \\
Current study - Quasi-static & Human ribs     & 13.5 $\pm$ 2.6 & 112.1 $\pm$ 24.5 & 1.06 $\pm$ 0.29\\
Current study - Dynamic      & Human ribs     &                & 124.6 $\pm$ 28.5 &                \\
\hline
\end{tabular}}
\caption{Summaries of the cortical bone material properties reported by various authors.}\label{tab:summary_mat_prop}
\end{table}

The analysis performed in the present study relies on the assumption that bone is homogeneous and isotropic, and therefore that the stress is greater in the gage area. However, the fracture location was not systematically in the gage area (table \ref{tab:mat_prop}), which suggest that the continuum mechanics approach has its limitations. This has several implications in terms of fracture prediction, as the microstructure - in addition to the geometry - alters the stress field in the cortical shell and therefore the fracture threshold and location. \citet{Li2010mep} performed an extensive computational analysis to determine which features of the rib geometry and structure needed to be included in a finite element model in addition to the adequate material properties to predict fracture under antero-posterior loading. Load and displacement at fracture, could be successfully predicted, whereas the fracture location predicted by the model did not match the experimental results. Both the cortical and trabecular bones were assumed elastic homogeneous and isotropic, and this may be why the fracture location was not correct, even with the cortical thickness distribution mapped from the actual ribs used in the experiments.

\subsection{Fracture prediction}
None of the published studies on bone coupons reported the fracture locations. However this would provide a valuable piece of information regarding the modeling approach that should be used to adequately predict bone fracture. As reported in the current study, fracture did not always took place in the gage area. Preliminary tests were performed on plastic homogeneous and isotropic plastic coupons prepared following the same procedure, and they all fractured in the gage area because of tension (the fracture line was perpendicular to the long axis of the coupons). Besides, the shear forces recorded with the three axis load cell were small compared to the load measured in the tensile direction (less than \unit{5}{\%}). This confirmed that the test apparatus itself did not generate artifactual fractures, and therefore the fracture characteristics are the results of the coupons material and structural properties. Therefore, although bone material might be adequately assumed to be homogeneous and elastic linear for sub-fracture behavior, the fracture properties seem to be dependent on the bone microstructure which is not properly accounted for with an elastic linear model. Researchers have had limited success to predict fracture location in finite element models \citep{Li2010mep}. Current finite element models of the ribs include the rib as structure composed of two distinct materials (trabecular and cortical bone). The results of the current study suggest that the delineation between cortical and trabecular bone has become less clear with improving imaging capabilities (figure \ref{fig:cross_section}): the cortical shell in the rib has pores, and the transition from the cortical to the trabecular bone can be somewhat blurred. Rather than using a two-part model for bone, a gradient approach may prove more accurate to predict the actual deformation modes and fracture locations under dynamic loading.
Rib bone material properties need to be refined to include a more precise model for fracture prediction, such as stress concentration caused by the presence of pores or by the connection between the trabeculae and the cortical shell. Micromechanics approaches such as the cohesive zone models \citep{Subit2009} would be worth evaluating as they can predict the onset of a crack based on the local microstructure.

\subsection{Experimental set-up}
The test apparatus designed for this study is promising: by design no load is applied when the coupon is affixed to the machine, and only tension is applied during loading. This led to a very high success rate during testing (no coupons were broken during handling or connection to the machine). However, it is not clear how the "loose" boundary conditions (compared to the traditional wedge clamps) affect the fracture outcome. When a crack initiates, the load distribution is the coupons changed suddenly and the clamps are likely to reorient themselves, which could lead to premature fracture. The same phenomenon could occur with plasticity. None of the data collected in the current study allow for checking whether the clamps accelerate the fracture process.

A non contact strain measurement system provided a non invasive method to estimate strain in the coupon. With the current method, only the average strain was estimated, similar to what a strain gauge would measure. This prevents from determining the link between the microstructure and the strain field, and therefore the fracture mechanism. Finally, the strain measurement procedure is 2D only, and consequently any relative motion of one of the clamps towards or away from the camera was seen as a change in strain, and therefore the measured strain could be overestimated (and the Young's modulus underestimated).

\section{Conclusion}
Predicting rib fracture based on the computational model of the thorax remains an elusive challenge: although the contribution of the geometry to the fracture mechanism has been demonstrated and included to some extent in finite models of the thorax \citep{Song2009,Li2011}, the contribution of the bone material properties has to be better described.  The intrinsic bone features that contribute to the onset of fracture are not well comprised by the commonly used material models (isotropic and homogeneous). The paradigm that describes bone as two entities --- trabecular or cortical --- needs to be revised to include a finer description of the bone microstructure. The results presented in this paper supplemented with past research highlight that the bone fracture mechanism is not well accounted for with a linear elastic model, at least for the rib bone. The attempt made in the current study to capture the bone microstructure by measuring the effective cross-sectional area along the coupons length proved unsuccessful. A more in-depth analysis of the bone microstructure (such as the direction of the voids and pores in the bone cortical shell) and how it modifies the strain field compared to the simplified approach that considers bone as elastic linear is required. It would allow to determine how the microstructure generates areas of weakness and strength to establish whether local strains or stresses play a role in the fracture mechanism.

\section{Acknowledgment}
The authors would like to acknowledge the contribution of James Bolton, Brian Overby amd Thomas Gochenour at the Center for Applied Biomechanics - University of Virginia for the design and fabrication of the test fixture, and of Stacy Hollins for the microstructure analysis, and thank Andrew Kemper and Stefan Duma from the Center for Injury Biomechanics (Virginia Tech - Wake Forest, USA) for sharing their expertise in bone coupon testing.

\section{Conflict of interest}
The authors declare that they do not have any conflict of interest in this study.

\appendix

\section{Cross-sectional images from computed micro-tomography}
\label{app1}

\begin{figure}[!h]
\begin{center}
  \centering
   \includegraphics[width=80mm]{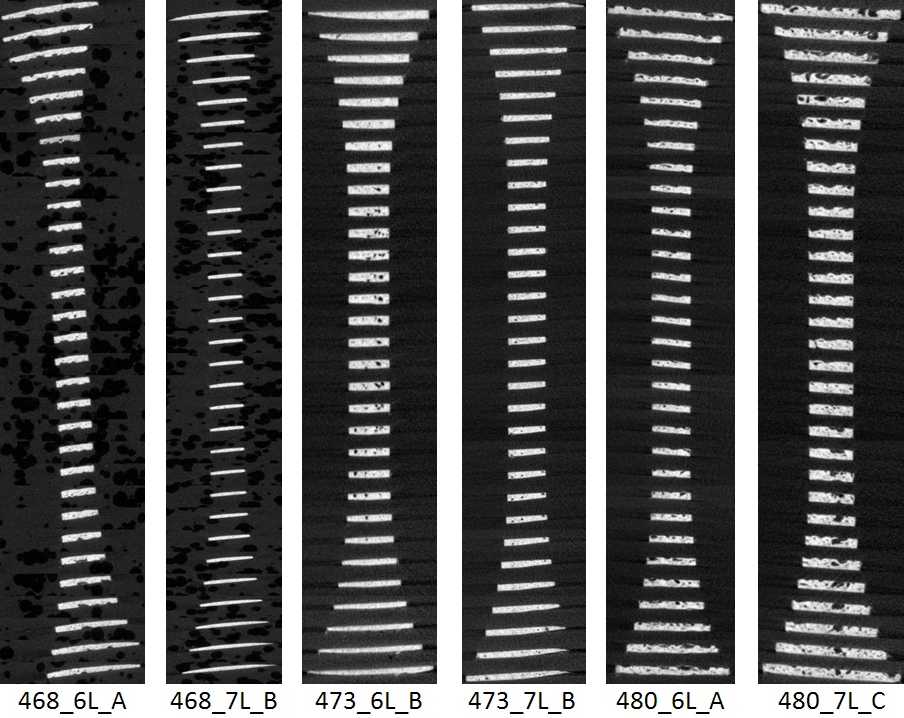}
    \caption{Images of the cross-section obtained from micro-computed tomography (coupons tested under quasi-static loading). Only one every ten images is included.}
  \end{center}
\end{figure}

\begin{figure}[!h]
\begin{center}
  \centering
   \includegraphics[width=50mm]{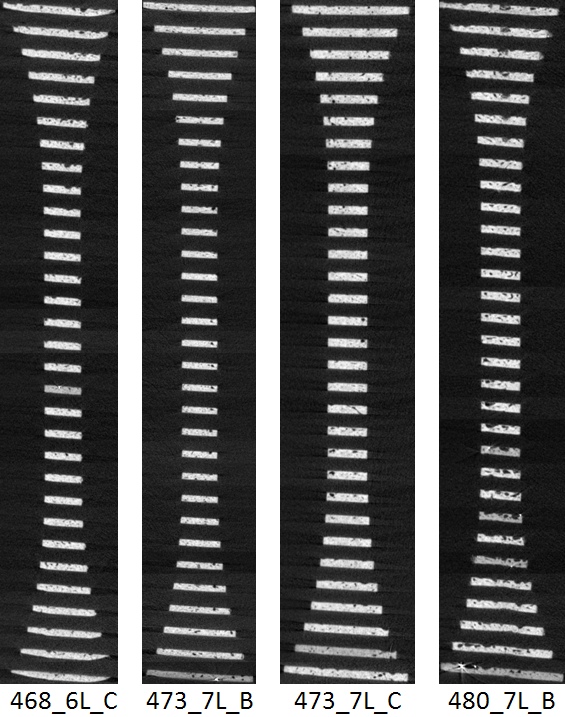}
    \caption{Images of the cross-section obtained from micro-computed tomography (coupons tested under dynamic loading). Only one every ten images is included.}
  \end{center}
\end{figure}

\FloatBarrier

\bibliographystyle{model2-names}
\bibliography{subit2011_coupon_bib}
\end{document}